\documentclass{b98proc}

\begin{document}

\title{EXCHANGE CURRENTS IN RADIATIVE HYPERON DECAYS AND HYPERON CHARGE RADII}
      
\author{GEORG WAGNER}
\address{Institut f\"ur Theoretische Physik, 
         Universit\"at T\"ubingen, 
         Auf der Morgenstelle 14,
         72076 T\"ubingen, Germany}

\maketitle

\abstracts{Radiative decays of decuplet hyperons and octet hyperon charge 
           radii are evaluated in a chiral constituent quark model
	   emphasizing the role of exchange currents. 
	   Exchange currents largely cancel for the M1 decay amplitudes, while 
	   they dominate the E2 amplitude. 
	   Due to the pseudoscalar meson cloud the charge radii of $\Sigma^-$
	   and $\Xi^-$ are almost as large as the proton radius, 
	   in agreement with recent experimental results from SELEX.
	   Strangeness suppression is weakened by exchange currents
	   for several observables.}  

\section{Introduction}
\label{sec:intro}
In this contribution we summarize our analysis of exchange current effects in 
the radiative hyperon decays\cite{Wag98a} and in octet hyperon charge 
radii\cite{Wag98b}.
With the advent of new facilities, detailed measurements of the radiative decays
of some $\Sigma^{\ast}$ and $\Xi^{\ast}$ hyperons\cite{Rus95} are under way.
Charge radii of some hyperons are now under experimental 
investigation\cite{Esc98}, too.
Many model calculations of hyperon decays (see Ref. list in\cite{Wag98a}) and
electromagnetic radii (see Ref. list in\cite{Wag98b}) are available 
for comparison.

The reasons for increased interest in these observables are twofold.
The $E2/M1$ ratios 
in the radiative hyperon decays contain information on deformations, if not of 
the valence quarks, then of the non-valence-quark distributions in the baryons.
Second, the observables are sensitive to SU$_F$(3) flavor symmetry breaking and
strangeness suppression.
Therefore, comparison of model predictions with experiment may pin down the 
relevant non-valence degrees of freedom in the effective quark-quark 
interaction.

\section{Exchange currents in the chiral constituent quark model}
\label{sec:model}
Constituent quarks emerge as the effective quasi-particle degrees of freedom 
in low-energy QCD due to the spontaneously broken chiral symmetry.
In our non-relativistic realisation of the chiral quark model a two-body 
confinement potential is used.
The pseudoscalar (PS) meson octet, that constitute the Goldstone bosons of the
symmetry breaking, provide the intermediate range interactions between quarks.
At short range, residual one-gluon exchange  is included.
 For the baryon wave functions we use spherical $(0s)^3$ oscillator states 
and SU$_{SF}$(6) spin-flavor states.
A detailed discussion of the Hamiltonian, parameters, wave functions,
hyperon masses and magnetic moments can be found in Ref.\cite{Wag95}.

\begin{figure}[htb]
  \psfig{figure=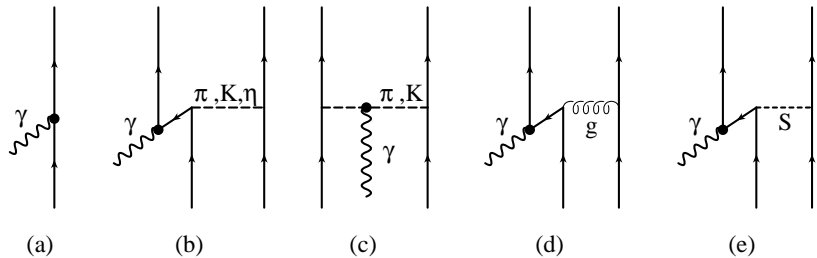,height=1.35in}
\caption{Electromagnetic one- and two-body currents as required by the 
         continuity equation.
         (a) Impulse approximation, (b) PS-meson pair current 
	 ($\pi , K, \eta$), (c) PS-meson in-flight current ($\pi , K$), 
	 (d) gluon-pair current, and (e) scalar exchange current 
	 (i.e. confinement).
\label{figure:currents}}
\end{figure}

Guided by the continuity equation, the electromagnetic currents to be considered
for the above described Hamiltonian are 
constructed\cite{Wag95,Buc91} by a non-relativistic reduction 
of the relevant Feynman diagrams shown in Figure \ref{figure:currents}.

\vspace{-0.1cm}

\vspace{-0.6cm}
\section{Radiative hyperon decays}
\label{sec:decay}
Table \ref{table:mamo} collects our results on the radiative
decays of decuplet hyperons\cite{Wag98a}.
Individual exchange current contributions to the $M1$ transition moments can be 
as large as 60$\%$ of the impulse approximation. 
However overall, like for the octet baryon magnetic moments\cite{Wag95}, 
exchange currents provide less than 10$\%$ corrections to the 
impulse approximation results.
Strangeness suppression in impulse approximation, i.e.\ in the 
valence quark picture, is considerable due to $m_u/m_s\simeq$ 0.6
(first column in Table \ref{table:mamo}).
Strangeness suppression is for all six strange decays
reduced when exchange currents are included.
The transition magnetic moments for the negatively charged hyperons
deviate considerably from the SU$_F$(3) flavor-symmetric value 0.

The transition quadrupole moments $Q$ shown in Table \ref{table:mamo} 
receive large contributions from the PS-meson and gluon-pair diagrams,
Figs.\ \ref{figure:currents}(b) and \ref{figure:currents}(d).
The $E2$  moments would be zero in impulse approximation 
for spherical valence quark configurations.
The dynamical origin for the allowed exchange current contributions is a double
spin-flip transition\cite{Buc91} of the two quarks involved.
The gluon contributes strongly to most transition quadrupole moments, on the
average $\sim$2/3 of the total $E2$ moment.
The experimental results may give important hints on the relevance of 
effective gluon degrees of freedom in hadron properties.  

The decay width $\Gamma \propto \vert A_{3/2} \vert^2 + \vert A_{1/2} \vert^2$
is related to the helicity amplitudes $A_{3/2}$ and $A_{1/2}$, which can be 
expressed as linear combinations of the $M1$ and $E2$ transition 
formfactors\cite{Buc91}.
The $E2/M1$ ratio of the transition amplitudes is commonly defined as 
$E2/M1 = \frac{\omega M_N}{6} Q/\mu$ with the resonance frequency $\omega$.
Due to cancellations of exchange current contributions to the $M1$ 
transition amplitude and the relative smallness of the $E2$ amplitude,
the decay widths $\Gamma$ are dominated by the $M1$ impulse approximation.
Only restricted informations on non-valence quark effects should 
be expected from the experiments here, 
similar to the situation for octet magnetic moments.

All model calculations yield large (the largest) $E2/M1$ ratios for 
negatively charged states.
The $E2/M1$ ratio for the decays of negatively charged hyperons are 
particularly  model dependent\cite{Wag98a} due to the smallness of both 
the $E2$ and $M1$ contributions.
We reiterate that in our calculation non-zero E2/M1 ratios are uniquely due to
non-valence degrees of freedom.

\vspace{-0.40cm}
\begin{table}[htb]
\caption[M1 Transition moments]{Transition magnetic 
        $\mu$ and quadrupole $Q$ moments of decuplet baryons.
        The impulse and exchange current contributions from the PS meson cloud
        and the gluon and scalar 
        quark-quark interactions to the transition magnetic moments
        are listed separately.
        The quadrupole PS-meson ${Q}_{\rm{PS}}$ and gluon-pair ${Q}_{\rm{G}}$ 
        contributions are given individually. 
        The last two columns contain the radiative decay widths 
        $\Gamma$ and $E2/M1$ ratios. 
\label{table:mamo}}
\begin{center}
  \footnotesize
\begin{tabular}{| l | r  r  r  r | r  r | r  r |} \hline 
\rule[-1mm]{0mm}{4.5mm} &
$\mu_{\rm{Imp}}$ & $\mu_{\rm{PS}}$ & $\mu_{\rm{G,Scal}}$ & $\mu_{\rm{Tot}}$ & 
${Q}_{\rm{PS}}$  & ${Q}_{\rm{G}}$  & $\Gamma$            & $E2/M1$          \\
& [$\mu_N$] & [$\mu_N$] & [$\mu_N$] & [$\mu_N$] 
& [fm$^2$]  & [fm$^2$]  & [keV]     & [$\%$]                 \\[0.05cm] \hline
$\gamma p\leftrightarrow \Delta^+$                    &
 2.828 &  .312 & -.609 & 2.533 & -.031 & -.058 & 350  & -3.65\\ 
$\gamma \Sigma^+ \leftrightarrow \Sigma^{\ast +}$     &
 2.404 &  .029 & -.166 & 2.267 & -.040 & -.051 & 105  & -2.9 \\
$\gamma n\leftrightarrow \Delta^0$                    &
 2.828 &  .312 & -.609 & 2.533 & -.031 & -.058 & 350  & -3.65\\ 
$\gamma \Sigma^0 \leftrightarrow \Sigma^{\ast 0}$     &
-0.990 & -.013 &  .078 &-0.924 &  .013 &  .016 & 17.4 & -2.3 \\
$\gamma \Lambda \leftrightarrow \Sigma^{\ast 0}$      &
 2.449 &  .154 & -.250 & 2.354 & -.007 & -.041 & 265  & -2.0 \\ 
$\gamma \Xi^0 \leftrightarrow \Xi^{\ast 0}$           &
 2.404 & -.020 &  .044 & 2.428 & -.004 & -.035 & 172  & -1.3 \\
$\gamma \Sigma^- \leftrightarrow \Sigma^{\ast -}$     &
-0.424 & -.004 &  .009 &-0.419 &  .014 &  .018 & 3.61 & -5.5 \\
$\gamma \Xi^- \leftrightarrow \Xi^{\ast -}$           &
-0.424 &  .009 & -.045 &-0.460 &  .004 &  .012 & 6.18 & -2.8 \\ \hline 
\end{tabular} 
\end{center}
\end{table}

\vspace{-0.6cm}
\section{Octet hyperon charge radii}
\label{sec:radii}

Constituent quarks are obtained through the dressing of current quarks by
their strong interactions.
 For an electromagnetic probe they appear as extended objects. 
The cloud of the constituent quark is dominated by $q\bar q$ pairs with 
pseudoscalar meson quantum numbers.
Guided by vector meson dominance, an electromagnetic constituent 
quark size can be assigned\cite{Wag98b}. 

Our results\cite{Wag98b} for the various one- and two-body
contributions to the charge radii are given in Table \ref{table:radii}.
The experimental proton and neutron charge radii are well described within our
model including gluon and PS-meson exchange currents. 

In impulse approximation, SU$_F$(3) symmetry breaking leads to a reduction of 
the charge radii with increasing strangeness content of the hyperon.
As a consequence the proton, $\Sigma^-$, and $\Xi^-$ have successively 
smaller charge radii. 
This strangeness suppression almost disappears when exchange currents are 
included.  
Due to the pseudoscalar meson cloud in the hyperons, the charge radius of the 
$\Sigma^-$ (or $\Xi^-$) turns out to be almost
as large as the proton charge radius.
A similar trend may be indicated by the recent experimental results from 
SELEX presented at this conference\cite{Esc98}.
SELEX obtain a $\Sigma^-$ charge radius of
$r^2_{\Sigma^-}$ = 0.60 $\pm$ 0.08 (stat) $\pm$ 0.08 (syst) fm$^2$, 
which agrees nicely with our result.

\vspace{-0.4cm}
\begin{table}[htb]
\caption[Charge Radii]{Charge radii of octet hyperons. 
        The PS-meson ($\pi ,K, \eta$) and the gluon-pair 
        ${r}_{\rm{G}}^2$ exchange currents are listed separately,
        as well as the scalar exchange current contributions. 
\label{table:radii}} 
  \vspace{0.1cm}
\begin{center}
  \footnotesize
\begin{tabular}{| l | c  c  c  c | c | c |} \hline
& \rule[-1mm]{0mm}{4.5mm}$r_{\rm{Imp}}^2$ & {$r_{\rm{PS}}^2$} & 
  {$r_{\rm{G}}^2$} & {$r_{\rm{Scal}}^2$} & {$r_{\rm{Tot}}^2$} &
  $\sqrt{\vert r_{\rm{Tot}}^2\vert }$ \\
&[fm$^2$] &[fm$^2$] &[fm$^2$] & [fm$^2$] & [fm$^2$] & [fm] \\[0.05cm] \hline  
$p$ &        0.736 &-0.057 & 0.123 &-0.150 & 0.651 & 0.807 \\
$\Sigma^-$ & 0.691 &-0.028 & 0.054 &-0.074 & 0.643 & 0.802 \\
$\Xi^-$    & 0.633 &-0.022 & 0.001 &-0.025 & 0.587 & 0.766 \\
$\Sigma^+$ & 0.861 &-0.029 & 0.118 &-0.125 & 0.826 & 0.909 \\ \hline
$n$ &         0    &-0.035 &-0.082 &   0   &-0.117 & 0.342 \\
$\Sigma^0$ & 0.085 & 0.000 & 0.032 &-0.028 & 0.088 & 0.296 \\
$\Lambda$  & 0.085 &-0.024 &-0.014 &-0.028 & 0.017 & 0.131 \\
$\Xi^0$    & 0.169 &-0.014 &-0.034 &-0.031 & 0.091 & 0.302 \\
\hline
\end{tabular} 
\end{center} 
\end{table} 

\vspace{-0.6cm}
\section*{Acknowledgments}
This work has been performed in collaboration with 
A.\ J.\ Buchmann and Amand Faessler (University of T\"{u}bingen, Germany).
I thank the DFG for a postdoctoral fellowship WA1147/1-1.

\def\Journal#1#2#3#4{{#1} {\bf #2}, #3 (#4)}
\def\NCA{\em Nuovo Cimento}
\def\NPA{{\em Nucl. Phys.} A}
\def\NPB{{\em Nucl. Phys.} B}
\def\PLA{{\em Phys. Lett.} A}
\def\PLB{{\em Phys. Lett.} B}
\def\PLD{{\em Phys. Lett.} D}
\def\PL{{\em Phys. Lett.}}
\def\PRL{\em Phys. Rev. Lett.}
\def\PREV{\em Phys. Rev.}
\def\PREP{\em Phys. Rep.}
\def\PRD{{\em Phys. Rev.} D}
\def\PRC{{\em Phys. Rev.} C}
\def\ZPC{{\em Z. Phys.} C}
\def\ZPA{{\em Z. Phys.} A}


\vspace{-0.4cm}
\begin{thebibliography}{99}
\bibitem{Wag98a} 
  G. Wagner, A. J. Buchmann, A. Faessler, \Journal{\PRC}{58}{1745}{1998}.
\bibitem{Wag98b} 
  G. Wagner, A. J. Buchmann, A. Faessler, 
            \Journal{\PRC}{58}{}{1998}, Dec. 1998. 
\bibitem{Rus95} 
  J. S. Russ,	    \Journal{\NPA}{585}{39c}{1995};
  R. A. Schumacher, \Journal{\NPA}{585}{63c}{1995}.
\bibitem{Esc98} 
  I. Eschrich for the SELEX Collaboration, these proceedings.
\bibitem{Wag95} 
  G. Wagner, A. J. Buchmann, A. Faessler, \Journal{\PLB}{359}{288}{1995}.
\bibitem{Buc91} 
  A. J. Buchmann, these proceedings;
  A. J. Buchmann, E. Hern\'{a}ndez, A. Faessler, \Journal{\PRC}{55}{448}{1997}; 
  A. J. Buchmann, E. Hern\'{a}ndez, K. Yazaki, \Journal{\PLB}{269}{35}{1991}.
\end{thebibliography}
\end{document}